# MICROCONTROLLER-DRIVEN MPPT SYSTEM FOR ENHANCED PHOTOVOLTAIC EFFICIENCY: AN EXPERIMENTAL APPROACH IN NEPAL

Diwakar Khadka[1], Satish Adhikari[1], Atit Pokharel[1], Sandeep Marasinee[1], Aayush Pathak[1]

[1]*Department of Electrical and Electronics Engineering*, *Kathmandu University, Dhulikhel Nepal*

*\*Corresponding author: idiwakarkhadka@gmail.com*

## ABSTRACT

Solar energy utilization in places like Dhulikhel, Nepal, is often obstructed by unpredicted environmental factors and existing technological barriers. The challenges encountered often result in fluctuating energy outputs, hindering the transition to greener energy solutions. To tackle these issues, this study introduces a custom-designed Maximum Power Point Tracking (MPPT) controller, seamlessly incorporated into a microcontroller-based battery charging system. This approach seeks to enhance the efficiency of photovoltaic (PV) systems, aligning with the global shift towards renewables. The research's primary objective is to enhance PV module power yield employing MPPT techniques, thereby reducing dependency on non-renewable energy sources. Key goals include real-time MPP tracking for optimal power extraction from PV modules and the integration of a real-time monitoring mechanism for PV and battery states. Leveraging a coordinated interplay of sensors measuring temperature, voltage, and current, vital metrics are fed to the microcontroller. This, in turn, generates a precise Pulse Width Modulation (PWM) signal, fine-tuning the voltage regulation of the buck-boost converter Metal Oxide Semiconductor Field Effect Transistor (MOSFET) for optimal operation. The adopted approach emphasizes monitoring environmental metrics, overseeing power outputs, and generating PWM signals to adeptly manage the buck-boost converter MOSFET voltage. The system also prioritizes balanced load alignment to boost power transfer and improve charging efficiency. An integrated LCD screen provides clear data visualization, allowing users to oversee and fine-tune the system's performance. Concurrently, data is transmitted hourly to the ThingSpeak cloud platform, facilitating real-time monitoring capabilities showcasing the potential of this system as a sophisticated IoT application. As a result of these integrations, an efficiency improvement of approximately 37.28% was observed. In essence, this research underscores the profound impact of merging advanced technologies within the renewable energy sector, offering a robust blueprint for enhancing energy stability and productivity. Building on the innovations and approaches introduced in this project, it's anticipated that it will set the stage for groundbreaking developments in renewable energy, guiding towards a greener and more sustainable future. Moreover, this study lays the groundwork for infrastructural advancements and encourage community participation in embracing green solutions, especially in regions similar to Dhulikhel.

## INTRODUCTION

Renewable energy plants, particularly solar power plants, are becoming more prevalent worldwide due to their eco-friendliness and the availability of solar radiation. Furthermore, the modular design of photovoltaic (PV) plants makes them versatile and adaptable for various power needs, making them a widely utilized solution in both industrial power generation and small-scale power supply system[1], [2]. The maximum efficiency achievable by a solar panel is limited to around 25%. Nevertheless, solar cells exhibit a commendable ability to generate power. Maximum Power Point Tracking (MPPT) is a technique used in photovoltaic (PV) systems to optimize the voltage and current levels. MPPT are employed to get the maximum output from a PV system [3], [4]. In a study conducted by Randriamanantenasoa et al., a comparative analysis of various Maximum Power Point Tracking (MPPT) techniques was undertaken. These techniques included Perturbation and Observation (P&O), Fuzzy Logic Controller (FLC), Artificial Neural Network & PI controller (ANN-PI), and Artificial Neural Network & Sliding Mode (ANN-SM). PV cell characteristics (I-V or V-P) are nonlinear and changes with insolation and temperature. The study found that Perturbation and Observation (P & O) was less suitable for managing dynamic changes in weather conditions. In contrast, the Artificial Neural Network (ANN) technique, particularly ANN-SM (Sliding Mode), demonstrated greater suitability for efficiently adapting to dynamic weather changes in the context of MPPT [5]. A recent study demonstrated that Model Reference Adaptive Control (MRAM) for MPPT surpasses the traditional Perturb and Observe (P&O) method in terms of efficiency, voltage and current ripple reduction, error rates, and convergence speed. The MRAM-based controller achieves an average tracking efficiency of 99.77% under varying temperatures and 99.69% in changing radiation conditions, while rapidly reaching the Maximum Power Point (MPP) in just milliseconds [6].

In the paper titled 'An Improved MPPT Control Strategy Based on Incremental Conductance' by Liqun Shang et al. emphasizes the importance of PV power generation in solar energy due to environmental concerns. Given the drawbacks of conventional methods, ensuring efficient MPPT is essential. These constraints are addressed in the research by introducing an improved incremental conductance technique that improves accuracy, speed, and stability. Its performance is rigorously assessed using MATLAB simulations, with an emphasis on monitoring efficiency and responsiveness to environmental changes. The paper also explores the function of institutional investors in PV systems. The simplicity and accuracy of this method increase tracking effectiveness and practical utility under a variety of circumstances [7].

This paper focuses on the design and implementation of a microcontroller-based battery charge controller with Maximum Power Point Tracker (MPPT) for photovoltaic (PV) power systems, aiming to improve efficiency and extend battery lifespan. The MPPT charge controller based on Incremental Conductance (IC) technique [8] is designed to operate the PV module at the peak power point, delivering maximum power to the batteries.

 The Constant Voltage MPPT method[9], which maintains a constant PV voltage output but is inefficient for considering the effect of temperature and solar irradiance. The Incremental Conductance Method involves calculating the rate of change of conductance with respect to voltage in a PV module, helping determine whether to increase or decrease the operating voltage to reach the maximum power point [10], [11].

A DC-DC Buck Boost converter [12] was designed to adjust the PV module voltage accordingly he powers generated by a solar panel can vary substantially during the day, regardless of whether Maximum Power Point Tracking (MPPT) technology is employed. These variations are primarily attributable to a range of factors, including alterations in solar conditions like sunlight angle, shading, weather, and the sun's position in the sky [13].

To assess how well MPPT technology performs, the power output from a solar panel was examined with and without MPPT for 24 hours in Dhulikhel, Nepal. The effectiveness of MPPT was observed in variations of weather conditions, including both sunny and cloudy days which were found to have a substantial impact on the performance of the solar MPPT system. The system operated at its highest efficiency during sunny days, generating a maximum output power of 36W. Conversely, its performance was considerably diminished during cloudy days, yielding a maximum output of 25W. The acquired data was transferred into a cloud-based storage system, enabling remote monitoring and analysis. This research paper lies in its contribution to the development of efficient solar energy systems in Nepal, which has a high potential for solar energy generation. The scope of the research paper is limited to the examination of the power output of a solar energy system with and without the use of MPPT and the impact of irradiance levels on the system's power output.

### a) MATERIALS AND METHODS

Maximum Power Point Tracking (MPPT) is essential for optimizing the efficiency of solar panels when connected to batteries or the utility grid. It ensures the solar panels operate within their appropriate voltage and current values to maximize energy extraction under varying conditions [14], [15]. MPPT solar charge controllers are crucial in solar systems with batteries, as they regulate voltage between solar panels and batteries, enhancing battery life, systematic protection and promotes longer lifespan of batteries. Moreover, grid-tied inverters also incorporate MPPT tracking system to extract maximum power from PV arrays.

In this study, a 40-watt solar panel was utilized which generates power depending on factors such as its efficiency factor, irradiance and temperature. The temperature sensor was utilized to measures ambient atmospheric temperature which directly affect the power output of panel. Likewise, panel output was monitored by voltage and current sensor. The data from those sensors were fetched to microcontroller; which then performed further processing, and transmitted a generated required PWM signal which was sent to the MOSFET gate pin of Buck-Boost converter for regulation of voltage level. Load impedance was adjusted to match the panel's impedance so that maximum power transfer was ensured. Harvested optimal power was supplied to a 12-volt battery with the adjusted voltage and current level. Battery voltage was sensed and fed back to the microcontroller which in turn utilize it to regulate the charging process. Information from various sensors, such as voltage, temperature, and current, along with battery status, maximum power point data, and current, was displayed on an LCD screen and also securely transmitted to the ThingSpeak cloud platform [16]. This cloud-based repository allows remote real-time monitoring and analysis exploring the IoT applications of such systems.

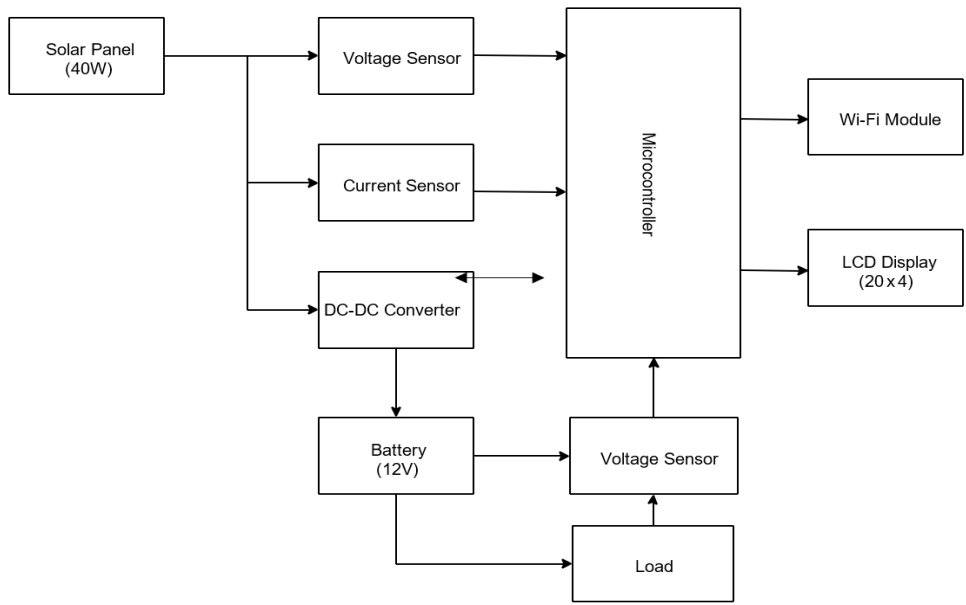

Figure 1: Block diagram of experimental setup

**2.1. MPPT Algorithm (Incremental Conductance Method)**

The incremental conductance (IC) method is a maximum power point tracking (MPPT) algorithm that tracks the maximum power point (MPP) of a photovoltaic (PV) array by comparing the incremental conductance (dI/dV) to the negative of the conductance (-I/V) [17].

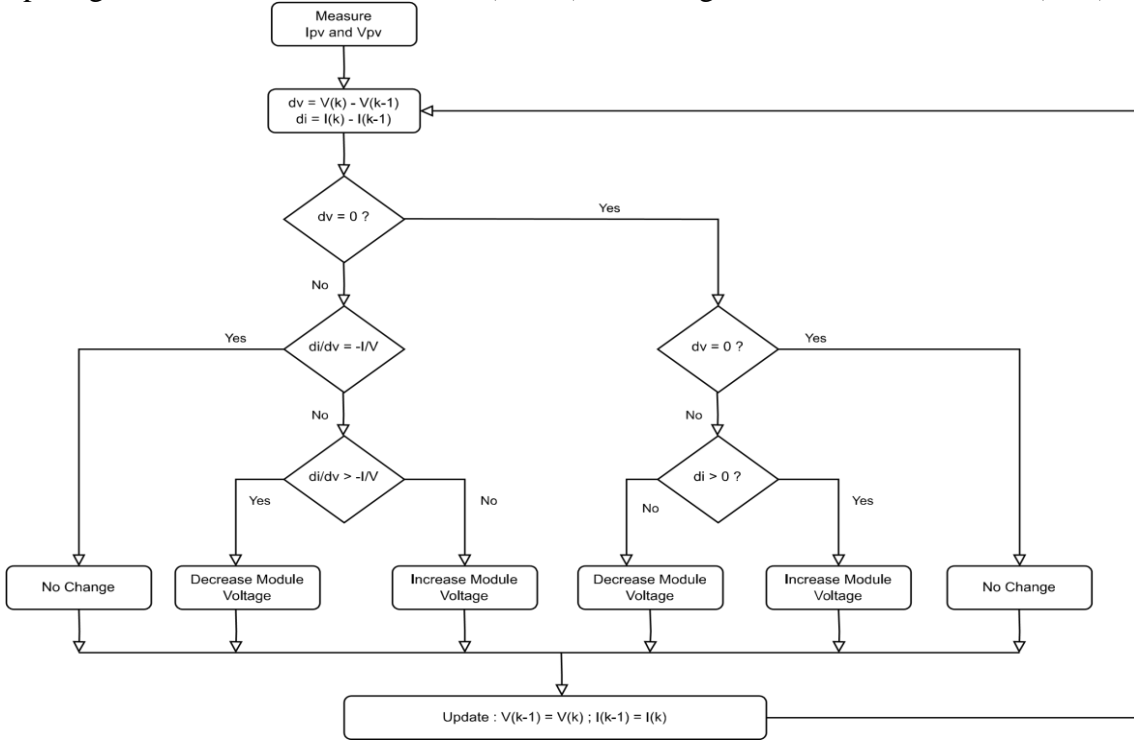

Figure 2: MPPT algorithm schematics

The IC method is based on the principle that the MPP is the point on the PV curve where the incremental conductance is equal to the negative of the conductance [18].

This can be expressed mathematically as follows:

$$\frac{I}{V} = -\frac{dI}{dV} \qquad (1)$$

The IC method works by continuously measuring the PV voltage and current and calculating the incremental conductance. If the incremental conductance is greater than the negative of the conductance, the MPPT controller increases the PV voltage. If the incremental conductance is less than the negative of the conductance, the MPPT controller decreases the PV voltage. This process continues until the MPP is reached [19].

## 2.2. Buck Converter

The buck converter, also known as a step-down converter, is a switching topology that transforms a DC input voltage, $V_{in}$, into a smaller DC output voltage, $V_{out}$. This conversion occurs in two phases: when the switch is on and when it's off. During the "on" phase, when the MOSFET or switch is active, it supplies current to the load. Initially, the current flow to the load is limited because energy is stored in the inductor (L) [20]. Consequently, the load current and the charge on the output capacitor ($C_{out}$) gradually increase during this "on" period.

Additionally, throughout this phase, the diode experiences a significant positive voltage, causing it to be reverse biased. When, the MOSFET or switch is turned off, the energy stored in the magnetic field around the inductor is released back into the circuit. The voltage across the inductor now reverses polarity compared to its voltage during the "on" period. Sufficient stored energy remains in the circuit to sustain current flow during this "off" period.

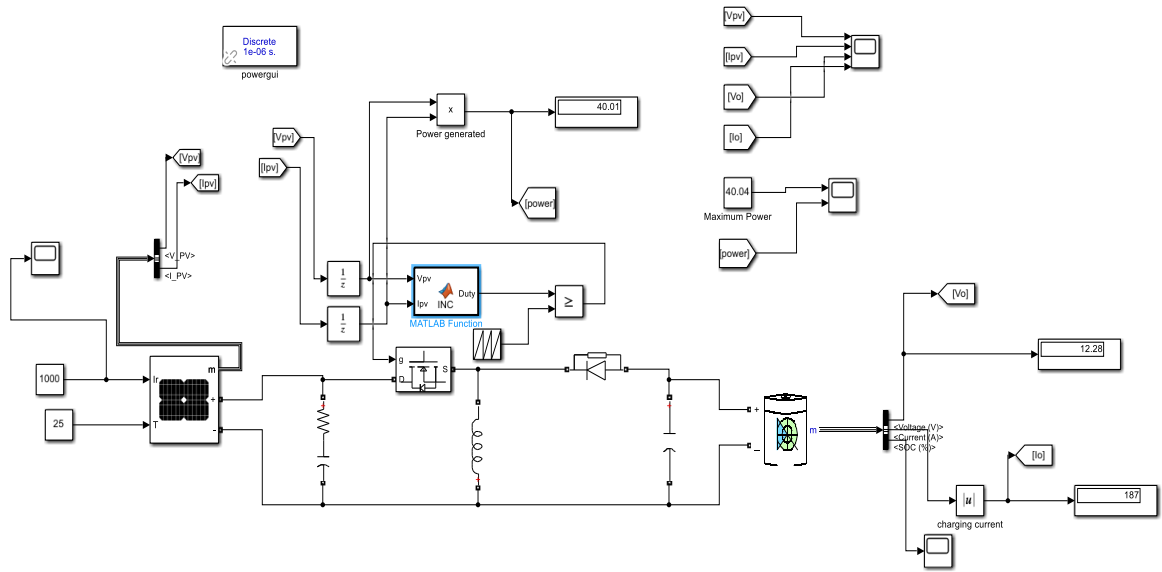

Figure 3: Schematics of buck converter with MPPT in Simulink

Calculation of Converter Design:
Maximum power of solar panel: 40w
Input voltage = 18.2v
Output voltage =12v
Switching frequency = 50Khz
$I_{ripple}$ = 20%
$V_{ripple}$ = 2%
Output Current =  = 3.33A
Current ripple = 10 % of 3.33A = 0.333A
Voltage ripple = 1% of 12V = 0.12V

$$L = \frac{Vop(Vip - Vop)}{FSW * Irp * Vrp} = 33H$$

$$C = \frac{Irp}{8 * FSW * Vrp} = 250\mu F$$

When a load is directly connected to a solar cell, the operating point of the panel seldom aligns with its peak power. The impedance encountered by the panel primarily dictates this operating point. Properly adjusting the impedance ensures that the panel operates at peak power.

Referring to the aforementioned circuit diagram, MPPT is realized in Simulink using the Incremental Conductance method. The primary inputs to the PV panel are irradiance and temperature, both of which significantly influence the panel's output. The Incremental Conductance method is employed to identify the optimal voltage required to achieve maximum power for a given irradiance level. The output of this algorithm, expressed in terms of duty cycle, is then relayed to the buck converters. This facilitates the charging of the battery at its maximum potential power.

**2.3. Interfacing the Solar Panel with a Load in the Absence of MPPT**

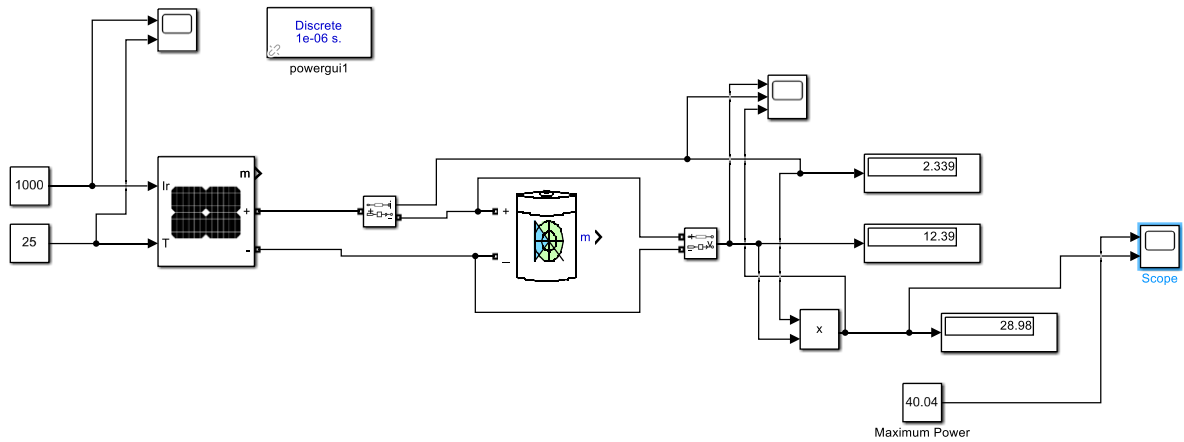

Figure 4: PV panel with load without MPPT

In this setup, the solar panel is directly connected to the battery without incorporating the MPPT circuit. Given that the input parameters of the solar panel, specifically irradiation and temperature, are held constant, the output from the solar panel remains consistent over the observed time frame.

## 2.4. Circuit simulation of the system

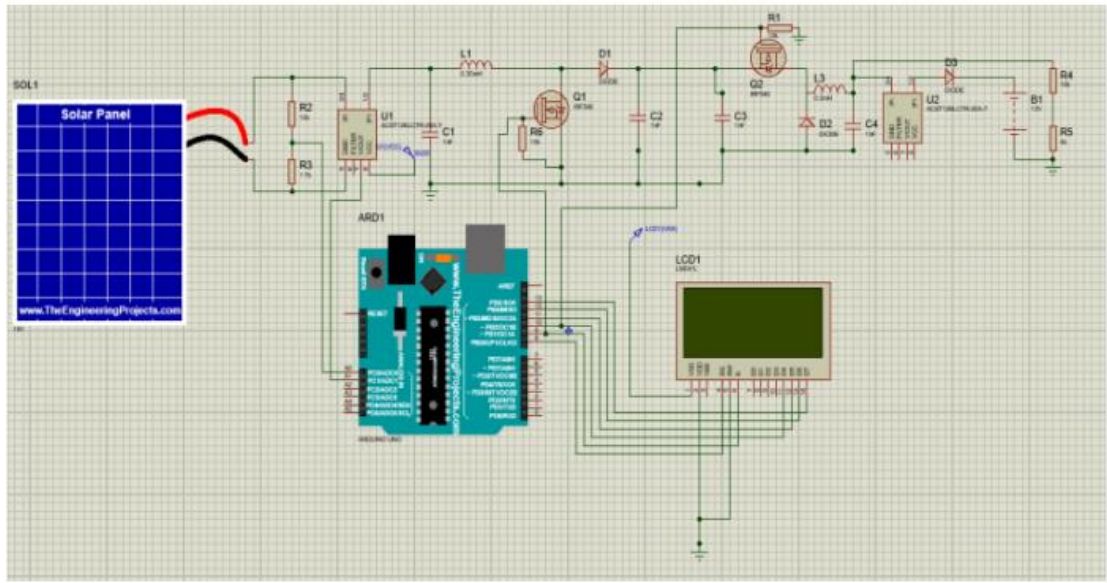

Figure 5: Circuit simulation on Proteus

The depicted circuit simulation, created in Proteus, represents our design blueprint. This simulation served as the foundation for our hardware implementation, ensuring a seamless transition from our digital design to the physical system.

## RESULT AND DISCUSSIONS

In this study, our primary objective was to assess and optimize the efficiency and output power of a photovoltaic (PV) system under two distinct scenarios: with and without the implementation of the Maximum Power Point Tracking (MPPT) circuit. Implementing the buck/boost converter alongside the incremental conductance method algorithm, we were able to simulate the behavior of the PV system under these conditions. Specifically, we plotted graphs for irradiance levels of 1000w/m$^2$ and 500w/m$^2$ in MATLAB for research purposes. Our findings, as depicted through these MATLAB-generated graphs as well as output generated by our hardware product, reveal a tangible disparity in the power generated by the PV panel in the presence versus the absence of MPPT. This section delves into a comprehensive analysis of these results and their implications on the potential efficiency gains associated with the use of MPPT in solar applications.

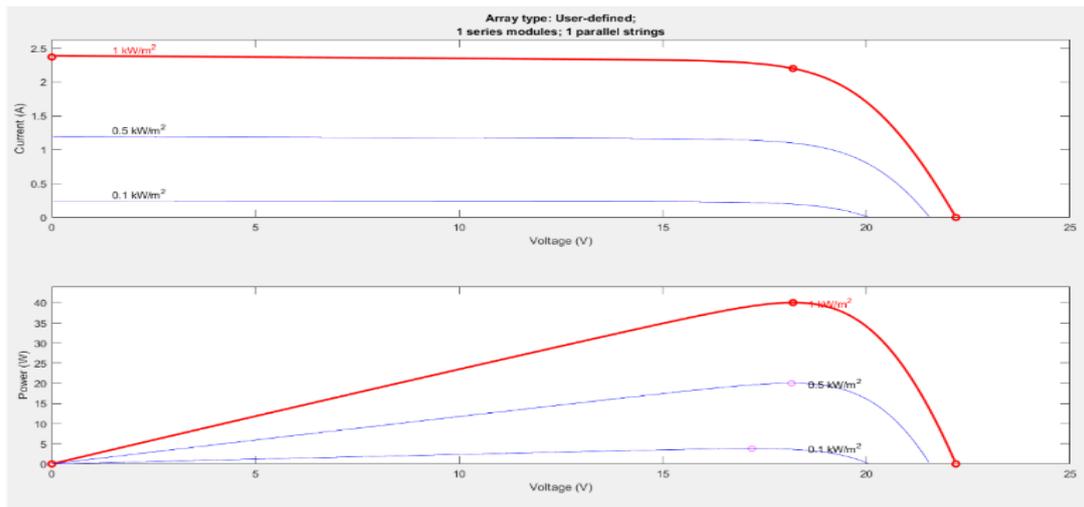

Figure 6: Variation of current and voltage with Irradiance

From the above-mentioned graph, it is evident that the maximum power achievable at an irradiance of 1000 w/m$^2$ is at a voltage of 18.2V. By employing the MPPT circuit, this maximum power voltage can be harnessed; however, without the MPPT, the battery charges at a suboptimal power level, reducing the system's overall competence.

**1.1. Comparison of the system with MPPT and without MPPT**

The system's power output is examined both with and without the use of MPPT. As illustrated in the subsequent figure, the expected maximum power at a specific irradiance level (i.e., 1000w/m^2) is 40W, represented by the black dotted line. The orange dotted line demonstrates the system's power output when the MPPT is not in place. With the incorporation of MPPT, there is a noticeable increase in the system's power output, as depicted by the red line in the plot, where the power incrementally rises to reach 40W. The power output of a solar energy system is intricately tied to the level of irradiance it receives, as depicted in the accompanying figure. Irradiance refers to the amount of radiant energy per unit area that reaches the Earth's surface.

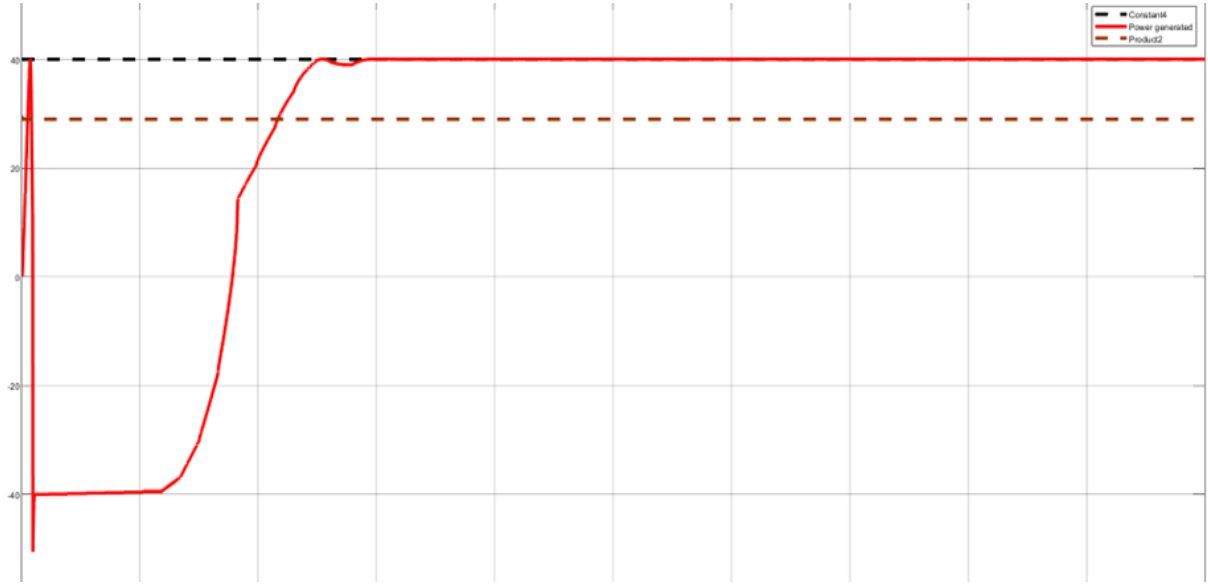

Figure 7: Power out comparison with and without MPPT

When the sun's rays are intense, leading to high irradiance levels, solar panels capture more energy and thus deliver a higher power output. Conversely, during periods of reduced irradiance, such as on cloudy days or at specific times of the day, the panels receive less radiant energy. This results in a decrease in the system's maximum power generation capacity. The output of the panel with varying irradiance is shown in the graph below.

**1.2. IV parameters with varying irradiance**

All the system voltages and current also varies according to the variation of the input factor of the solar panel. System output realized with the help of MATLAB Simulink is shown below:

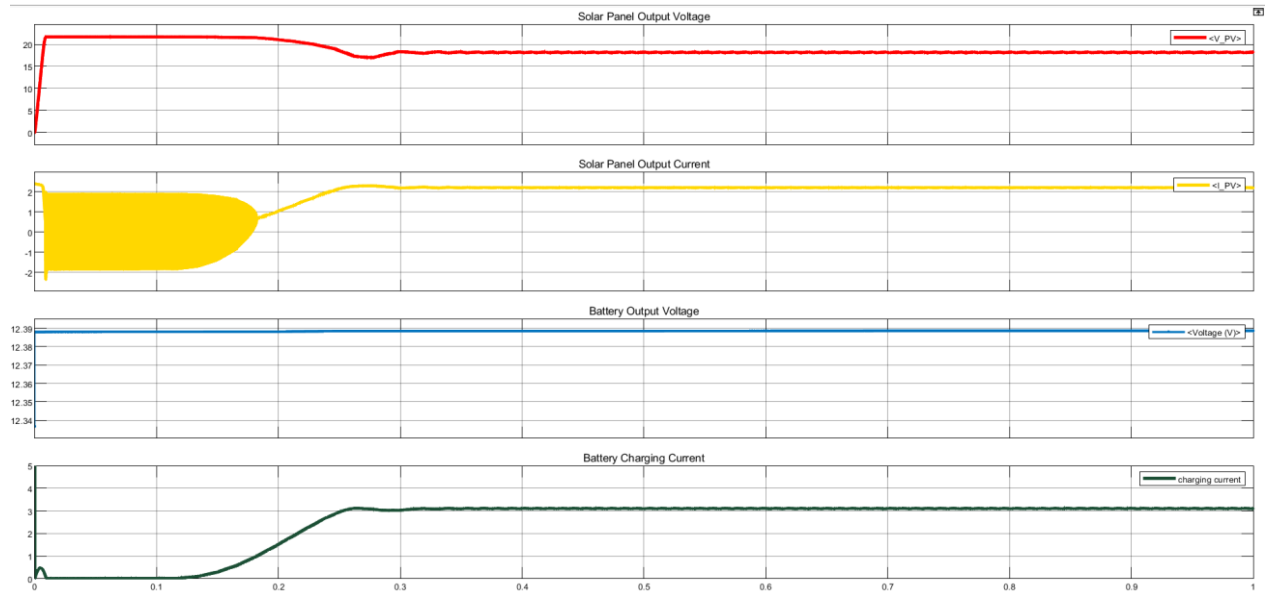

Figure 8: System output with irradiance 1000w/m2

The output current of the solar panel is significantly influenced by irradiance levels. As irradiance increases, the solar panel's output current rises, enabling a greater power supply to charge the battery. Conversely, when irradiance decreases, the solar panel's output current diminishes, resulting in a reduced current available for battery charging.

Referring to the figure above, with an irradiance of 1000w/m^2, the solar panel's output current stands at approximately 2.1A, while the battery's charging current is 3A.

Now if the irradiance is decreased to 500w/m^2, both the output current of the PV panel and the battery's charging current will see a corresponding reduction. Solar output current is reduced to 1A and the battery charging current is reduced to 1.5A when the irradiance is decreased from 1000w/m2 to 500 w/m2 shown in the graph below.

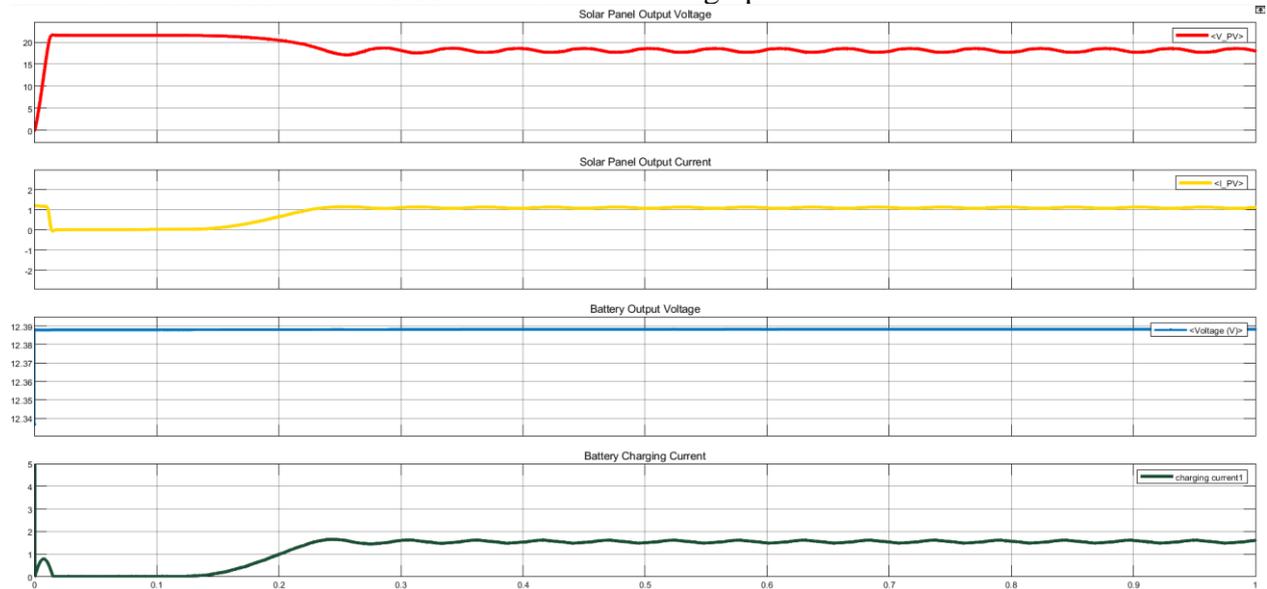

Figure 9: System output with irradiance 500 w/m2

Now if the charging current of the battery is high battery will charge at higher speed and if the charging current is low battery will charge at lower speed. The behavior of the battery charging is also analyzed using the scope block of Simulink

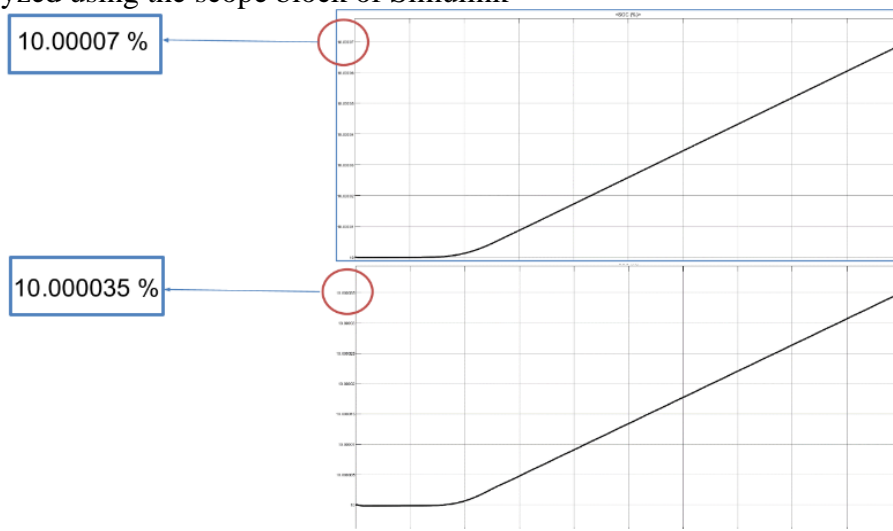

Figure 10: Comparison of battery charging varying irradiance

From the Figure [10], it is evident that with an irradiance of 1000 w/m^2, the battery's charge increases from 10% to 10.00007% in just 1 second. Similarly, when the irradiance drops to 500 w/m^2, the charge rises from 10% to 10.000035% in the same duration. If the charging current remains constant at 3A, the battery will be fully charged in 10 hours. However, with a reduced current of 1.5A, the charging duration extends to 20 hours, given the battery's capacity of 30Ah.

### 1.3. Hardware Prototype

Below, we present our hardware implementation, a tangible display of our research. The system had been designed using a buck/boost converter, an LCD display, a MOSFET driver, and an Arduino to orchestrate the interactions. This hardware setup has not only allowed us to validate our initial simulations but has also been instrumental in gathering the crucial data presented in subsequent sections to strengthen our research findings.

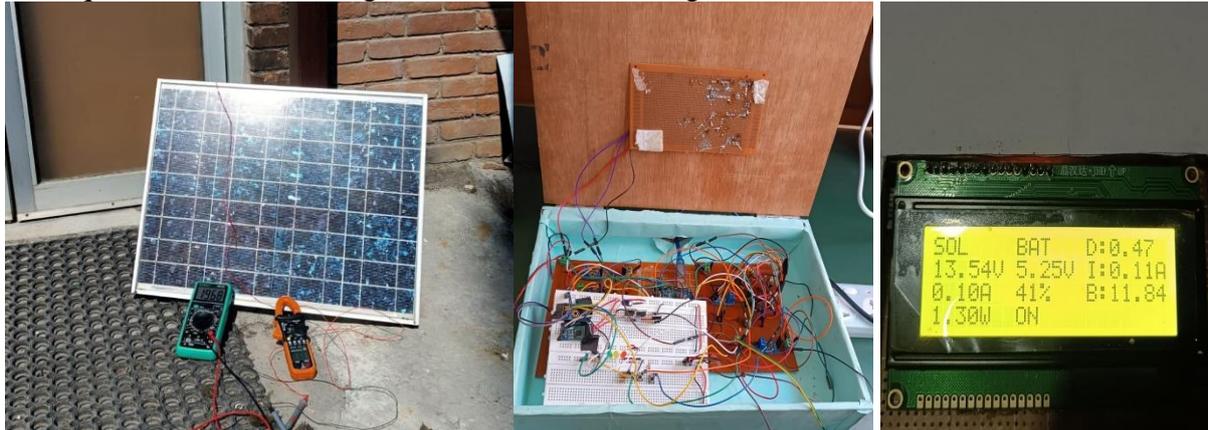

Figure 11: Hardware implementation and LCD display for real-time monitoring

The LCD display interface, Figure [11] shows a comprehensive data of Photo Voltaic (PV) arrays data including values of current, voltage and power generation. Likewise, it provides detailed insights into battery system, including its charging voltage, state of charge and whether it was presently activated. Moreover, screen provides critical information regarding the duty cycle, battery status voltage and current level.

### 1.4. Power output from the hardware designed

Table [1] shows the hourly recorded power levels yielded by the PV panel in both the scenarios including with MPPT integration and without MPPT integration in a sunny day in Dhulikhel, Nepal.

Table 1: Output of solar panel in a sunny day

| Time | Without MPPT Power Generation (W) | With MPPT Power Generation (W) |
|---|---|---|
| 6:00 AM | 14 | 25 |
| 7:00 AM | 15 | 26 |
| 8:00 AM | 18 | 28 |
| 9:00 AM | 20 | 28 |
| 10:00 AM | 22 | 30 |
| 11:00 AM | 25 | 32 |
| 12:00 PM | 28 | 35 |
| 1:00 PM | 29 | 36 |
| 2:00 PM | 27 | 33 |
| 3:00 PM | 25 | 32 |
| 4:00 PM | 24 | 31 |
| 5:00 PM | 22 | 30 |
| 6:00 PM | 18 | 28 |

Figure [12] illustrates power generation from a solar panel system with and without MPPT technology over the course of a sunny day. Power production gradually rises throughout the day in both scenarios (with and without MPPT), reaching its highest point at 1:00 PM with 36W for MPPT and 29W without MPPT. Across all hours, MPPT consistently outperforms the non-MPPT system, demonstrating its efficacy in enhancing solar panel power output under changing sunlight conditions.

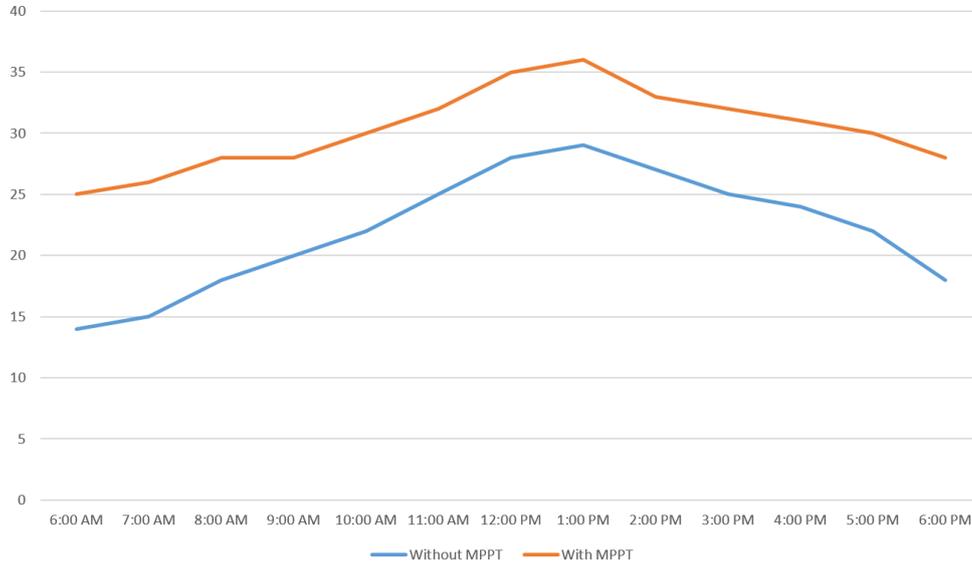

Figure 12: PV Solar Panel output during sunny day

Table [2] shows the hourly recorded power levels yielded by the PV panel in both the scenarios including with MPPT integration and without MPPT integration in a cloudy day.

Table 2: Output of solar panel in a cloudy day

| Time | Without MPPT Power Generation (W) | With MPPT Power Generation (W) |
|---|---|---|
| 6:00 AM | 9 | 15 |
| 7:00 AM | 10 | 16 |
| 8:00 AM | 11 | 16 |
| 9:00 AM | 13 | 18 |
| 10:00 AM | 12 | 16 |
| 11:00 AM | 15 | 20 |
| 12:00 PM | 17 | 22 |
| 1:00 PM | 18 | 25 |
| 2:00 PM | 16 | 23 |
| 3:00 PM | 15 | 19 |
| 4:00 PM | 13 | 18 |
| 5:00 PM | 13 | 18 |
| 6:00 PM | 10 | 15 |

Figure [13] illustrates power generation from a solar panel system with and without MPPT technology over the course of a cloudy day. Power production gradually rises throughout the day in both scenarios (with and without MPPT), reaching its highest point at 1:00 PM with 25

W for MPPT and 18W without MPPT. Across all hours, MPPT consistently outperforms the non-MPPT system. Although it outperforms the non MPPT system its output is significantly low compared to that of a Sunny Day with the MPPT.

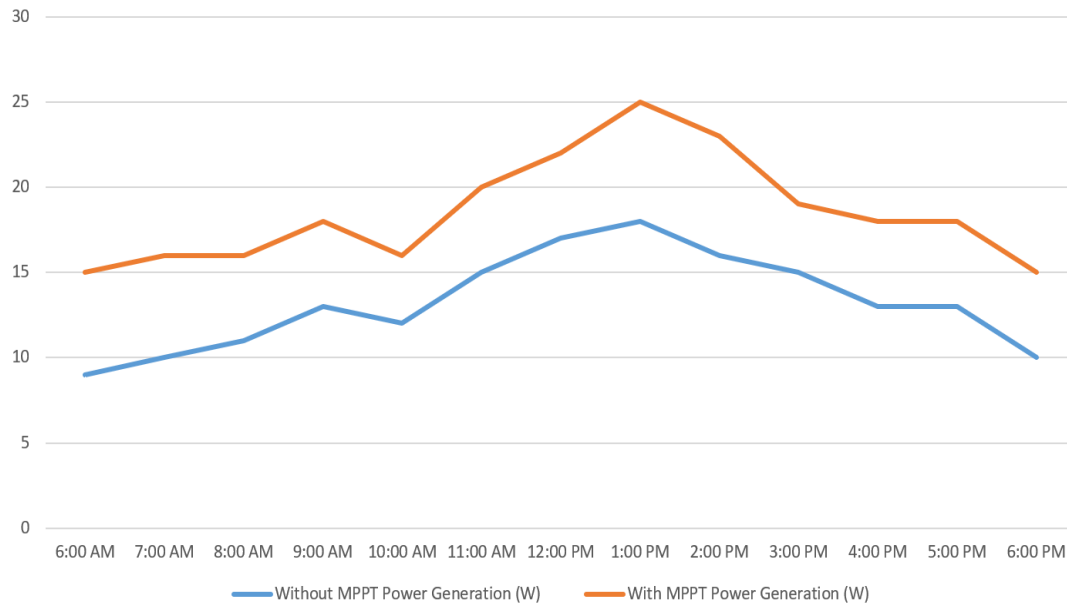

Figure 13: PV Solar Panel output during cloudy day

### 1.4.1. Comparison of Power Output in between Sunny and Cloudy Day

The table offers a comparison of solar power output with and without the use of Maximum Power Point Tracking (MPPT) at various times throughout the day. Power generation is measured in watts (W). The "Time" column lists specific hours, spanning from 6:00 AM to 6:00 PM. Subsequent columns showcase the power generation values under two scenarios: without MPPT and with MPPT. The column titled "Without MPPT Power Generation (W)" presents values representing the solar power output when MPPT is not employed. Conversely, the "With MPPT Power Generation (W)" column depicts the power output when the system incorporates MPPT technology. MPPT enhances solar panel power generation by continuously locating and utilizing the maximum power point, ensuring optimal panel efficiency. A side-by-side comparison of values in these columns highlights the tangible benefits of MPPT throughout the day. MPPT technology proves to be highly effective in boosting power generation on sunny days, achieving a peak output of 36W in the early afternoon, a notable improvement compared to the 28W maximum without MPPT. Similarly, on cloudy days, MPPT demonstrates its advantages, elevating power production to 25W at 1:00 PM, in contrast to the 18W peak seen without MPPT. These results highlight the efficacy of Maximum Power Point Tracking in optimizing solar panel performance, allowing them to operate nearer to their maximum power potential, particularly under less favorable weather conditions. Generally, power output with MPPT consistently exceeds that without MPPT, highlighting the efficacy of MPPT in boosting solar panel performance.

## CONCLUSION

In conclusion, this research addresses pivotal challenges in the adoption of solar energy, particularly in regions like Dhulikhel, Nepal, where environmental and technological barriers often limit its efficacy. Our study introduces a robust Maximum Power Point Tracking (MPPT) controller integrated into a microcontroller-based battery charging system, aiming to significantly enhance the efficiency of photovoltaic (PV) modules. Employing a sophisticated interplay of sensors and real-time tracking also with the exploration of possible IoT application, the study observed a notable improvement in efficiency by approximately 37.28%. This work not only serves as a blueprint for elevating energy stability and productivity but also advocates for a greener and more sustainable future. The study's findings extend beyond mere energy generation, stimulating infrastructural development and galvanizing community engagement in renewable energy initiatives. By melding advanced technologies and rigorous methodology, we contribute to the burgeoning field of renewable energy, pointing the way for future innovations.